# Spin Dynamics near the Superconductor-to-Insulator Transition in Impurity-Doped $YBa_2Cu_4O_8$

Yutaka ITOH, Takato MACHI, Nobuaki WATANABE, Seiji ADACHI and Naoki KOSHIZUKA

*Superconductivity Research Laboratory, International Superconductivity Technology Center, 10-13 Shinonome 1-chome, Koto-ku, Tokyo 135-0062*

We studied low-frequency spin dynamics near the impurity-induced superconductor-to-insulator transition for underdoped high-$T_c$ superconductor $YBa_2(Cu_{1-x}M_x)_4O_8$ (M=Ni, Zn) using the Cu nuclear quadrupole resonance (NQR) spin-echo technique. We observed remarkable suppression of the normal-state pseudo spin-gap and a loss of Cu NQR spectrum intensity at low temperatures around the critical impurity concentration.

Superconductor-to-insulator (or semiconductor) transition in a two-dimensional system is an intriguing phenomenon.[1] Heavily doped impurities induce the superconductor-to-insulator transition in underdoped high-$T_c$ superconductors at low temperatures.[2] Neutron scattering studies have revealed Zn-induced low-frequency spin fluctuations around the antiferromagnetic wave vector for underdoped $YBa_2(Cu_{1-x}Zn_x)O_{6+}$ and for Zn-doped $La_{2-x}Sr_xCuO_4$.[3] The dynamics of the impurity-induced superconductor-to-insulator transition, however, has not been sufficiently understood. In this Letter, we report the Cu NQR study on low-frequency spin dynamics around the *rf* frequency region (~30 MHz) for impurity-doped high-$T_c$ superconductor $YBa_2(Cu_{1-x}M_x)_4O_8$ (M=Ni, Zn) with the superconductor-to-insulator transition. The critical impurity concentration $x_c$ sufficient to destroy superconductivity is estimated to be about 0.04 for magnetic Ni ($3d^8$, $S=1$) and about 0.03 for nonmagnetic Zn ($3d^{10}$, $S=0$) in our samples. $YBa_2Cu_4O_8$ is a stoichiometric underdoped compound without an appreciable oxygen deficiency. Near the superconductor-to-insulator transition, we found that the impurities rapidly suppress the pseudo spin-gap behavior of the plane-site Cu(2) nuclear spin-lattice relaxation rate $1/T_1T$, although a loss of Cu(2) NQR spectrum intensity (wipeout effect) prevents precise measurement at low temperatures. Here, we associate the above $T_c$ behavior of Cu $1/T_1T$, which exhibits a maximum at a specific temperature and subsequently decreases with decreasing temperature, with the pseudo spin-gap.

Samples of powdered YBa$_2$(Cu$_{1-x}$M$_x$)$_4$O$_8$ (M=Ni, x=0, 0.009, 0.018, 0.030, 0.042; M=Zn, x=0.005, 0.010, 0.022) were synthesized by a solid-state reaction method with a high-oxygen-pressure technique.[4,5] The impurity content $x$ was estimated by inductively coupled plasma atomic emission spectroscopy, which is different from the nominal values in refs. 4-7. The superconducting transitions were observed at $T_c$=82 K for pure x=0, at $T_c$=66, 44, 15 K, <4.2 K for Ni doping of x=0.009, 0.018, 0.030, 0.042 and at $T_c$=68, 56, 27 K for Zn doping of x=0.005, 0.010, 0.022, from $dc$ magnetization measurements.

A coherent-type pulsed spectrometer was utilized for the Cu NQR measurements. The Cu NQR frequency spectra with quadrature detection were obtained by integration of the spin-echoes as the frequency was scanned point by point. Nuclear spin-lattice relaxation curves were measured by an inversion recovery spin-echo technique, where the $^{63}$Cu(2) nuclear spin-echo intensity $M(t)$ was recorded as a function of time interval $t$ after an inversion pulse, in a -$t$- /2- - echo sequence.

We analyzed the experimental recovery curve $p(t)$ 1-$M(t)/M($ ) by the exponential function times a stretched exponential function,[8]

$$p(t) = p(0)\exp[-\frac{3t}{T_{1\,HOST}} - \sqrt{\frac{3t}{\tau_1}}].$$
(1)

The fit parameters are $p(0)$, $(T_1)_{HOST}$ and $\tau_1$. $(T_1)_{HOST}$ is the Cu nuclear spin-lattice relaxation time due to the host Cu electron spin fluctuation. $\tau_1$ is the impurity-induced nuclear spin-lattice relaxation time. Equation (1) is an asymptotic form derived from the extended function with a direct wipeout effect.[8] As shown in ref. 7, however, the extended form gives an overestimation of the wipeout number $N_c$ above $T_c$, so that eq. (1) is employed under the next best policy.[9] Practically, it is hard to distinguish the respective contributions from various (direct or indirect) interactions between the Cu nuclei spin $I$ and the impurity spin $S$ within the Cu NQR technique. At least, a longitudinal direct dipole interaction ( $S_zI_\pm/r^3$, $r$ is a distance from an impurity to a nuclear) must yield the stretched exponential part of eq. (1).[8] The origin of $\tau_1$ for Ni is, at least, the direct dipole coupling with Ni impurity. Although the origin of $\tau_1$ for Zn is not clear within the framework of the present study, the Zn-induced neighboring local moments[10,11] or staggered moments[12] are promising candidates. According to ref. 13, the stretched exponential approximation of eq. (1) is applicable for higher impurity concentration.

Figure 1 shows $^{63,65}$Cu NQR frequency spectra of the plane-site Cu(2) (a) [(c)] and of the chain-site Cu(1) (b) [(d)] for the Ni[Zn]-doped YBa$_2$Cu$_4$O$_8$ at 4.2 K. Both Ni and Zn shift slightly and broaden the spectra. With the impurity doping, the peak frequency $^{63}\nu$(2) of $^{63}$Cu(2) NQR spectrum shifts to a higher frequency, whereas $^{63}\nu$(1) of $^{63}$Cu(1) shifts to a lower frequency, i.e. $^{63}\nu(2)/x$~6.7 MHz, $^{63}\nu(1)/x$~-6.7 MHz for Ni doping of $x$=0.03, and $^{63}\nu(2)/x$ ~4.6 MHz, $^{63}\nu(1)/x$~-2.3 MHz for Zn doping of $x$=0.022. These shifts are in parallel directions with those under the hydrostatic physical pressure (up to 3 GPa).[14] Thus, the observed shifts are attributable to the chemical pressure effect of impurity. The full-width at half maximum $^{63}$ (1, 2) of the $^{63}$Cu(1, 2) spectrum increases as the impurity increases, i.e. $^{63}$ (2)/$x$~26 MHz, $^{63}$ (1)/$x$ ~15 MHz for Ni doping of $x$=0.03, and $^{63}$ (2)/$x$~20 MHz, $^{63}$ (1)/$x$ ~10 MHz for Zn doping of $x$=0.022. The broadening may be due to local deformation of the electric field gradient around crystalline disorders or due to charge density corrugation around impurities, but not due to the static local ordering of staggered magnetization. The significant broadening of the Cu(1) NQR spectra for Ni doping supports the assumption that some amount of the impurity substitutes for the chain site, especially for Ni. In contrast to Ni, the Zn doping of $x$=0.022 in Fig. 1(c) induces the broad satellite lines around 28.8 MHz [26.6 MHz] for $^{63}$Cu(2) [$^{65}$Cu(2)], which have already been reported above 160 K in ref. 15. These Cu NQR satellite lines may be associated with the $^{89}$Y NMR satellite lines in ref. 11.

Figure 2 shows Ni[Zn]-doping dependence of the $^{63}$Cu(2) recovery curve at $T$=90 K (a) [(c)] and at 200 K (b) [(d)]. We measured $T$-dependence of the recovery curve at each peak frequency of the $^{63}$Cu(2) NQR spectrum in Figs. 1(a) and 1(c). The solid curves are fitted results based on eq. (1). For Ni doping, the fittings are satisfactory. But for Zn doping, some deviations at longer times $t$ are seen at high temperatures. The analysis at longer time $t$ has been discussed in a separate paper in ref. 7.

Figure 3 shows Ni[Zn]-doping dependence of the host relaxation rate $(1/T_1T)_{HOST}$ (a) [(c)] and the impurity-induced relaxation rate $1/\tau_1$ (b) [d]. The results for Ni doping of $x$=0.042 and for Zn doping of $x$=0.022 are quite new, and the other results are reproduced from refs. 6 and 16, in order to observe the change. The inset figure in Fig. 3(b) [in Fig. 3(d)] shows $T$ dependence of the integrated intensity of Cu(2) NQR spectrum times temperature ($I\times T$) for Ni doping of $x$=0.042 [for Zn doping of $x$=0.022], which is corrected by the Boltzmann factor ($1/T$) and extrapolated by the spin-echo decay curve with fitting a Gaussian times exponential function ($T_2$ correction). In Figs. 3(b) and 3(d), $1/\tau_1$ increases with Ni and Zn doping, because the relaxation center of impurity increases.

In Fig. 3(a), for Ni doping of $x$=0-0.03 ($T_c$=82-15 K), the pseudo spin-gap behavior of $(1/T_1T)_{HOST}$ is quite robust.[6] However, for Ni doping of $x$=0.042 ($T_c$ < 4.2 K), $(1/T_1T)_{HOST}$ changes into a Curie-Weiss behavior above about 60 K, which could not be simply explained by suppression of the broad peak of $(1/T_1T)_{HOST}$. The pseudo spin-gap behavior must be below 60 K. Although we cannot demonstrate the detailed change of $(1/T_1T)_{HOST}$ from the pseudo spin-gap behavior to the Curie-Weiss behavior at present, one may regard the change of $(1/T_1T)_{HOST}$ from $x$=0.03 to $x$=0.042 as the rapid suppression of the pseudo spin-gap behavior in the metallic regime[2, 5] above about 60 K. Below about 60 K, the wipeout effect as shown in the inset of Fig. 3(b) prevents the precise estimation of $(T_1)_{HOST}$ and $\tau_1$. At 4.2 K, the Cu(2) NQR signals have disappeared from around 30 MHz. A qualitatively similar change is seen for Zn doping of $x$=0.022 in Fig. 3(c), although in contrast to Ni, Zn suppresses largely the magnitude of $(1/T_1T)_{HOST}$ above about 100 K.[16] The wipeout effect on the Cu(2) NQR spectrum is demonstrated in the inset of Fig. 3(d), where the spectrum does not completely disappear at 4.2 K. Thus, we observed the rapid suppression of the pseudo spin-gap behavior and the low-temperature wipeout effect near the critical impurity concentration $x_c$.

We associate this wipeout effect with the localization effect in conduction at low temperatures.[17] The localization effect increases markedly the dwell time of itinerant electrons at a nuclear site, so that the correlation time $\tau(\ 1/T_{1,\ 2})$ of the local hyperfine field fluctuation becomes much longer than that in metallic conduction. Thus, slow spin fluctuations wipe out the Cu NQR signals from localized sites by causing rapid nuclear spin relaxation at low temperatures. The impurity-induced staggered magnetic moments[12, 18] must also contribute to the rapid relaxation, although neither a long-range magnetic order nor a spin glass transition[19] has been explored in the present materials.

One should be careful with eq. (1) in the wipeout effect below about 60 K. At high temperatures where the conduction is metallic,[2, 4, 5] the separation in the relaxation process between the host Cu spin-fluctuation and the impurity spin is clear in eq. (1). But, at low temperatures where the conduction is semiconductive[2, 5] and the wipeout effect works, the separation between the localized host electron spin and the impurity spin is not clear. Then, the physical meanings of $T_1$ and $\tau_1$ in the wipeout effect at low temperatures may have to be changed from those at high temperatures.

Figure 4 shows a magnetic phase diagram of the impurity effect, where the superconducting transition temperature $T_c$, the pseudo spin-gap temperature $T_s$ [the temperature of the maximum $^{63}(1/T_1T)_{HOST}$], and the pseudogap temperature $T_g(= E_g/k_B$, the pseudogap energy $E_g$ estimated from $^{89}$Y NMR Knight shift in refs. 20 and 21) are plotted against Ni or Zn content $x$. On approaching the superconductor-to-insulator transition, the pseudogap is robust,[11, 20-22] whereas the pseudo spin-gap is suppressed rapidly. The impurity effect on $T_s$ is different from the monotonic decrease of $T_s$ in ref. 23. In contrast with the long wavelength (uniform) mode, the short wavelength (antiferromagnetic) mode probed by $(T_1)_{HOST}$ is sensitive to the critical doping. Such a different response to impurity between the antiferromagnetic and the uniform modes is a familiar phenomenon in low dimensional magnetic insulators.[24, 25] Although the antiferromagnetic mode could be a critical soft mode on the boundary of a gapless region, the wipeout effect does not allow us to determine whether a quantum critical behavior is realized. Also, further quantitative discussion requires knowledge of the actual concentration of the in-plane impurity.

In conclusion, the pseudo spin-gap behavior of the $^{63}$Cu(2) nuclear spin-lattice relaxation rate is suppressed rapidly near the impurity-induced superconductor-to-insulator transition in the underdoped YBa$_2$Cu$_4$O$_8$. This is in contrast to the robust pseudogap behavior of the uniform spin susceptibility.[11, 20-22] The wipeout effect is also observed at low temperatures near the critical impurity concentration $x_c$. The existence of a long-range magnetic ordering or of a spin glass state in the wipeout effect remain topics for future discussion.

We thank Dr. T. Masui for fruitful discussions. This work was supported by New Energy and Industrial Technology Development Organization (NEDO) as Collaborative Research and Development of Fundamental Technologies for Superconductivity Applications.

**References**
1) M. P. A. Fisher, G. Grinstein and S. M. Girvin: Phys. Rev. Lett. **64** (1990) 587.
2) Y. Fukuzumi, K. Mizuhashi, K. Takenaka and S. Uchida: Phys. Rev. Lett. **76** (1996) 684.
3) K. Kakurai, S. Shamoto, T. Kiyokura, M. Sato, J. M. Tranquada and G. Shirane: Phys. Rev. B **48** (1993) 3485; H. Harashina, S. Shamoto, T. Kiyokura, M. Sato, K. Kakurai and G. Shirane: J. Phys. Soc. Jpn. **62** (1993) 4009; K. Hirota, K. Yamada, I. Tanaka and H. Kojima: Physica B **241** (1998) 817.
4) T. Miyatake, K. Yamaguchi, T. Takata, N. Koshizuka and S. Tanaka: Phys. Rev. B **44** (1991) 10139.
5) N. Watanabe, N. Koshizuka, N. Seiji and H. Yamauchi: Physica C **234** (1994) 361.
6) Y. Itoh, T. Machi, N. Watanabe and N. Koshizuka: J. Phys. Soc. Jpn. **68** (1999) 2914; Physica B **281**&**282** (2000) 914.
7) Y. Itoh, T. Machi, N. Watanabe and N. Koshizuka: submitted to Phys. Rev. B.
8) M. R. McHenry, B. G. Silbernagel, and J. H. Wernick: Phys. Rev. Lett. **27**, (1971) 426, Phys. Rev. B **5** (1972) 2958.
9) Y. Itoh, T. Machi, N. Watanabe and N. Koshizuka: to be published in Physica C (in press).
10) R. E. Walstedt, R. F. Bell, L. F. Schneemeyer, J. V. Waszczak, W. W. Warren, Jr., R. Dupree and A. Gencten: Phys. Rev. B **48** (1993) 10646.
11) A. V. Mahajan, H. Alloul, G. Collin and J.-F. Marucco: Phys. Rev. Lett. **72** (1994) 3100.
12) J.-H. Julien, T. Fehér, M. Horvatic, C. Berthier, O. N. Bakharev, P. Ségransan, G. Collin and J.-F. Marucco: Phys. Rev. Lett. **84** (2000) 3422.
13) P. Thayamballi and D. Hone: Phys. Rev. B **21** (1980) 1766.
14) T. Machi, M. Kosuge, N. Koshizuka and H. Yasuoka: *Advances in Superconductivity VII*, eds by K. Yamafuji and T. Morishita (Springer-Verlag, Tokyo, 1994) p. 151.
15) G. V. M. Williams, R. Dupree and J. L. Tallon: Phys. Rev. B **60** (1999) 1360.
16) Y. Itoh, T. Machi and N. Koshizuka: *Advances in Superconductivity XII*, eds by T. Yamashita and K. Tanabe (Springer-Verlag, Tokyo, 2000) p. 284.
17) S. E. Fuller, E. M. Meintjes and W. W. Warren, Jr.: Phys. Rev. Lett. **76** (1996) 2806.
18) J. Bobroff, H. Alloul, Y. Yoshinari, A. Keren, P. Mendels, N. Blanchard, G. Collin and J.-F. Marucco: Phys. Rev. Lett. **79** (1997) 2117.
19) J.-H. Julien *et al.*: cond-mat/0010362.
20) G. V. M. Williams, J. L. Tallon, R. Meinhold and A. Jánossy: Phys. Rev. B **51** (1995) 16503.
21) G. V. M. Williams, J. L. Tallon, R. Dupree and R. Michalak: Phys. Rev. B **54** (1996) 9532.
22) The robust pseudogap for critical impurity doping has been reported for the other high-$T_c$ cuprates. For Zn-doped YBa$_2$Cu$_3$O$_{6.64}$, see [A. V. Mahajan, H. Alloul, G. Collin and J. F. Marucco: Eur. Phys. J. B **13** (2000) 457]. For impurity-doped La$_{2-x}$Sr$_x$CuO$_4$, see [N. Ishikawa, N. Kuroda, H. Ikeda and R. Yoshizaki: Physica C **203** (1992) 284].

23) G.-q. Zheng, T. Odaguchi, Y. Kitaoka, K. Asayama, Y. Kodama, K. Mizuhashi and S. Uchida: Physica C **263** (1996) 367; G.-q. Zheng, T. Odaguchi, T. Mito, Y. Kitaoka, K. Asayama and Y. Kodama: J. Phys. Soc. Jpn. **62** (1993) 2591.
24) K. Uchinokura, T. Ino, I. Terasaki and I. Tsukada: Physica B **205** (1995) 234.
25) P. Carretta, A. Rigamonti and R. Sala: Phys. Rev. B **55** (1997) 3734.

**Figure Captions**

Fig. 1   Zero-field $^{63, 65}$Cu NQR frequency spectra of the plane-site Cu(2) (a) [(c)] and of the chain-site Cu(1) (b) [(d)] for Ni doping of $x=0$, 0.018, 0.030 [for Zn doping of $x=0$, 0.010, 0.022] at $T=4.2$ K. The solid curves are guides for the eye. Either impurity shifts slightly and broadens both NQR spectra of Cu(2) and of Cu(1). The Zn doping of $x=0.022$ induces the broad satellite lines denoted by the arrows, which agree with the result above 160 K in ref. 15.

Fig. 2   Ni[Zn]-doping effects on the planar $^{63}$Cu(2) nuclear spin-lattice recovery curves at 90 K (a) [(c)] and at 200 K (b) [(d)], at each peak frequency of the $^{63}$Cu NQR spectrum in Fig. 1(a) and 1(c). The solid curves are the least-squares fitting results based on eq. (1).

Fig. 3   Ni[Zn]-doping effects on the temperature dependences of the host $(1/T_1T)_{HOST}$ (a) [(c)] and of the impurity-induced $1/\tau_1$ (b) [(d)]. The solid curves are fitted results based on a Curie-Weiss law $C/(T+\Theta)$ ($C=2.3\times10^3$ s$^{-1}$ and $\Theta=200$ K for Ni doping of $x=0.042$, and $C=9.6\times10^2$ s$^{-1}$ and $\Theta=84$ K for Zn doping of $x=0.022$). With the heavy doping of impurities, $(1/T_1T)_{HOST}$ changes from the pseudo spin-gap behavior to the Curie-Weiss-like behavior. Inset figures in Fig. 3(b) and 3(d) show the temperature dependences of the integrated Cu NQR intensity times temperature ($I\times T$).

Fig. 4   Magnetic phase diagram drawn from the results in Fig. 3. Here, $T_c$ is the superconducting transition temperature, $T_s$ is the maximum temperature of $(1/T_1T)_{HOST}$ (the pseudo spin-gap temperature), and $T_g$ is the pseudogap temperature estimated from the $^{89}$Y NMR Knight shift in refs. 20 and 21. The solid and the dashed curves are guides for the eye. The "localization" region is inferred from the resistivity in refs. 2, 4 and 5.

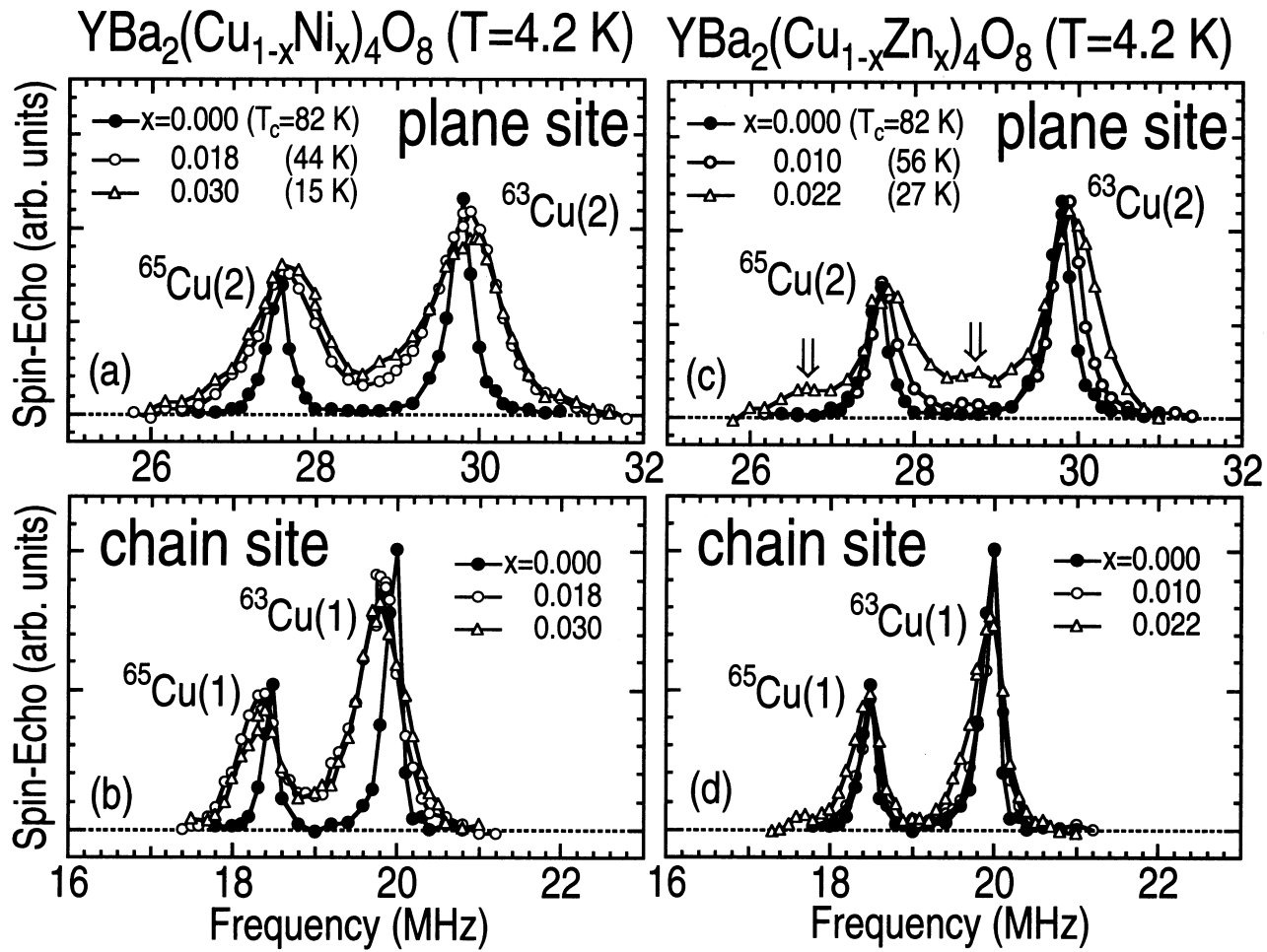

Fig. 1  Y. Itoh et al.

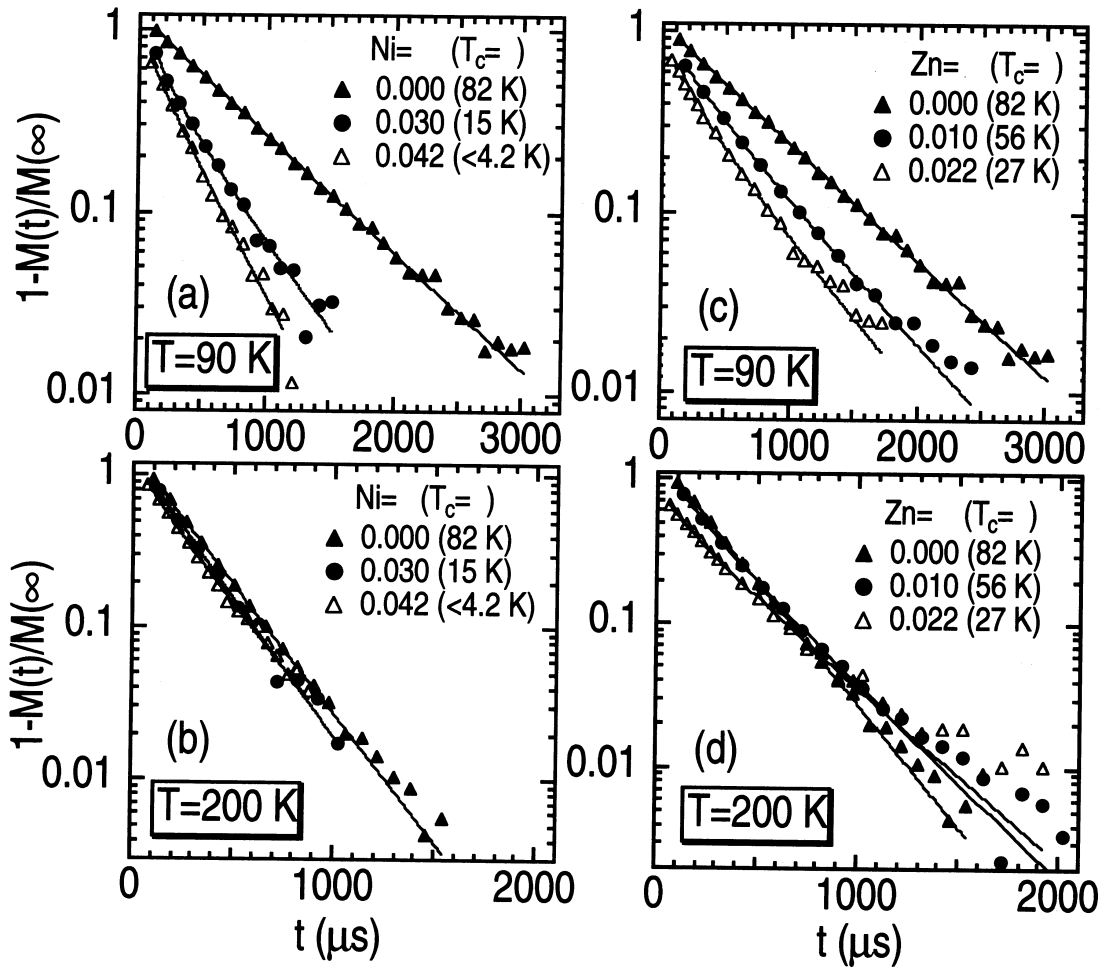

Fig. 2   Y. Itoh et al.

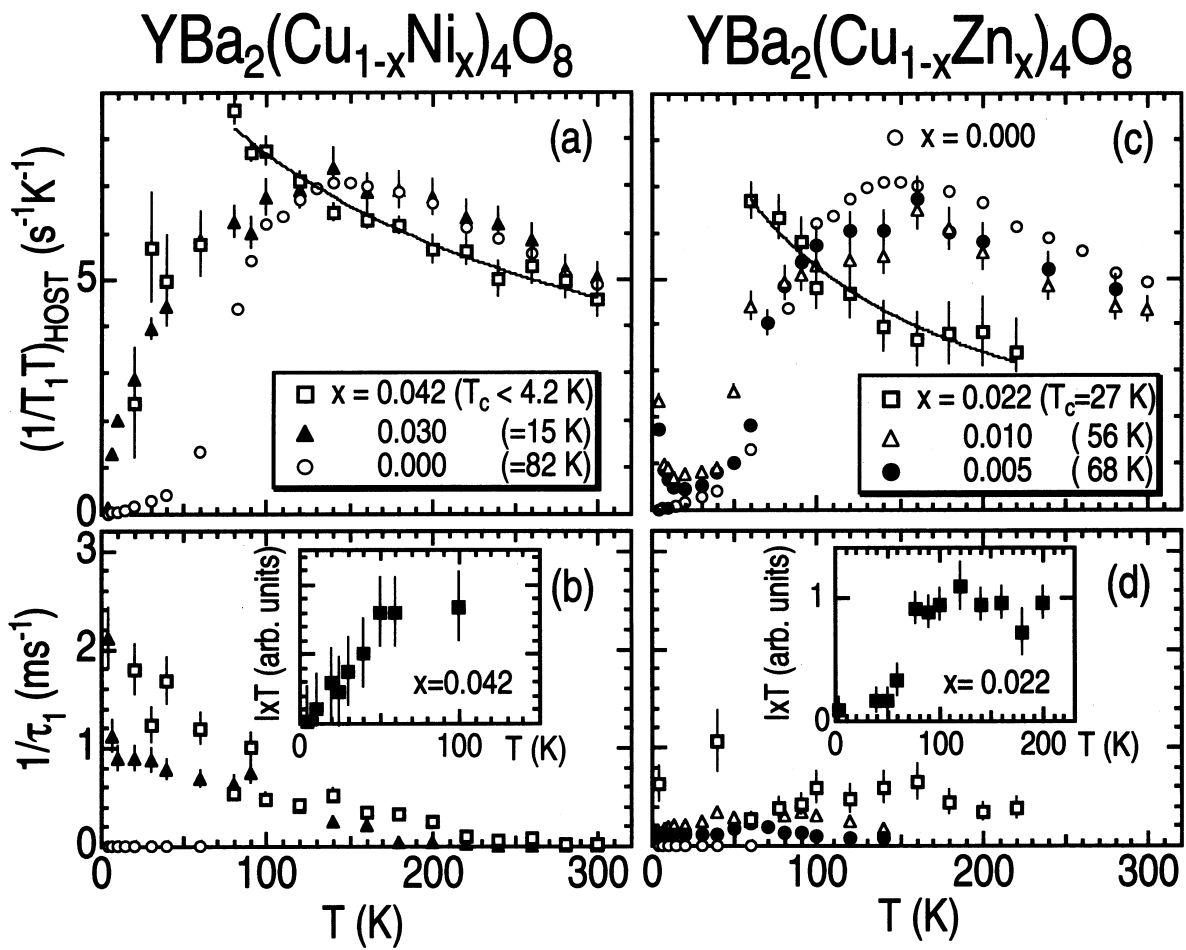

Fig. 3   Y. Itoh et al.

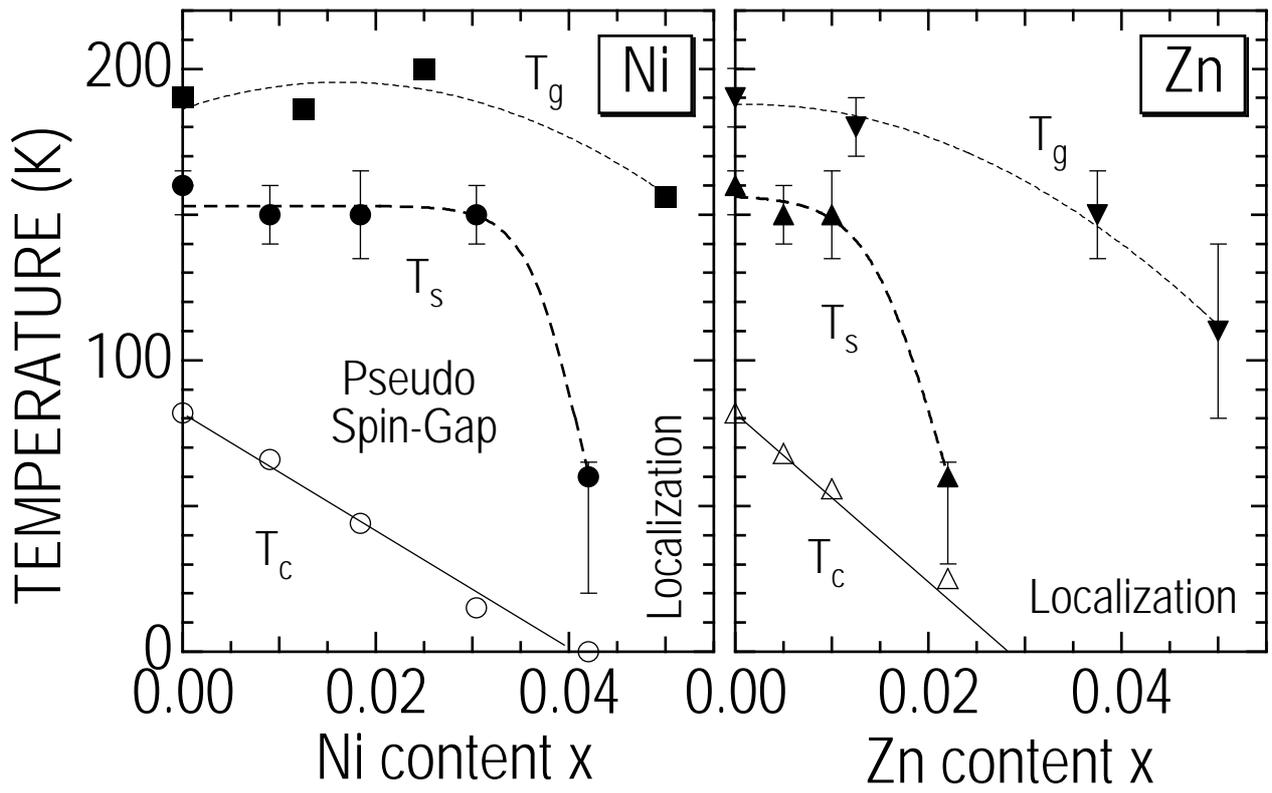

Fig. 4    Y. Itoh et al.